# Orbital competition of $Mn^{3+}$ and $V^{3+}$ ions in $Mn_{1+x}V_{2-x}O_4$

J. L. Jiao[1], H. P. Zhang[2], Q. Huang[3], W. Wang[1], G. Wang[1], Q. Ren[1], R. Sinclair[3], G.T. Lin[1], A. Huq[4], H. Cao[4], H. D. Zhou[3], M. Z. Li[2] and J. Ma[1,*]

[1]Key Laboratory of Artificial Structures and Quantum Control, School of Physics and Astronomy, Shanghai Jiao Tong University, Shanghai 200240, China

[2]Department of Physics, Beijing Key Laboratory of Opto-electronic Functional Materials & Micro-nano Devices, Renmin University of China, Beijing 100872, China

[3]Department of Physics and Astronomy, University of Tennessee, Knoxville, Tennessee 37996, USA

[4]Neutron Scattering Division, Oak Ridge National Laboratory, Oak Ridge, Tennessee 37381, USA

**Abstract**

The structure and magnetic properties of $Mn_{1+x}V_{2-x}O_4$ ($0 < x \leq 1$), a complex frustrated system, were investigated by heat capacity, magnetization, x-ray diffraction and neutron diffraction measurements. For $x \leq 0.3$, a cubic-to-tetragonal (c > a) phase transition was observed. For $x > 0.3$, the system maintained the tetragonal lattice. Collinear and noncollinear magnetic transition was also observed for all compositions. To reveal the dynamics of the ground state, first principle simulation was applied to not only analyze the orbital effects of $Mn^{2+}$, $Mn^{3+}$, and $V^{3+}$ ions, but also the related exchange energies.

## 1. Introduction

The geometry frustrated magnets compounds have been attracted a lot of attention because of the exotic magnetic properties at the degenerate ground state [1-3]. Typically, the geometrically frustrated lattice includes the corner- and side-shared triangle as the two-dimensional (2D) kagome and triangular compounds, and three-dimensional (3D) spinel and pyrochlore compounds, respectively [2, 4, 5]. Spinel oxide, $AB_2O_4$, is a natural geometric frustrated material on the pyrochlore sublattice at B site.

In this system, the spin, orbital, charge and lattice degrees of freedom are strongly correlated and compete with each other to induce varieties of ordered states, which influences the macroscopic properties, such as structural, electric, and magnetic properties [6-10]. Different physical properties have been observed in spinel oxides, including the colossal magnetoresistance (CMR) effect, giant magnetoelectric effect, and multiferroicity [11, 12]. In addition, the electrons on degenerate orbital could interact with crystal field and cause lattice distortion if they don't fill up the energy levels, which was known as Jahn-Teller (JT) effect [13]. According to the theory of Kugel and Khomskii [14], the orbital could couple the spin to relieve the magnetic frustration, resulting in a magnetic ordering state.

Vanadate spinel, $AV_2O_4$, has the magnetic $V^{3+}$ ($3d^2$, $S=1$) ions at the B-site, and is an ideal system to study the complex interplay of electron, spin, lattice and orbit. The octahedral crystal field accompanied by 6 $O^{2-}$-ions around the centered $V^{3+}$ ion can split the $3d$ orbitals of $V^{3+}$ ion into $t_{2g}$ orbitals with lower energy and $e_g$ orbitals with higher energy. Due to the Jahn-Teller effect, the degeneracy of $t_{2g}$ will be further released and two electrons of $V^{3+}$-ion will choose single orbit with lowest energy spontaneously. Several theoretical models have been proposed to clarify such orbital ordering of $V^{3+}$ ion [15-17].

In $MnV_2O_4$, the $V^{3+}$ ions are influenced under the internal magnetic field from the A-site $Mn^{2+}$ ions, which makes $MnV_2O_4$ exhibit several phase transitions: paramagnetism (PM)-to-collinear ferrimagnetism (CF) and collinear-to-noncollinear ferrimagnetism (NCF) along with a structural phase transition from cubic to tetragonal structure (c < $a$) [18, 19]. Since the magnetic and structural phase transitions happen at the same temperature, there is a debate on the causality of ferrimagnetic order and the tetragonal structural transition. Based on the previous work [19], the orbital effect of $V^{3+}$-ion was believed to relate to both magnetic and structural transitions. However, Ma et. al [20] found that the CF-NCF transition was decoupled from the cubic-tetragonal structural phase transition with substituted $Mn^{2+}$-ions by $Co^{2+}$-ions on A-sites as $Mn_{1-x}Co_xV_2O_4$, and the orbital ordering of $V^{3+}$-ion completely disappeared while the NCF phase still existed. These manifested the independence between Yafet and Kittel (YK) magnetic transition and structural phase transition. Although the $Co^{2+}$-ions only introduced an internal magnetic field as $Mn^{2+}$-ions, the itineracy increased in the system and the localized orbital ordering of $V^{3+}$-ions was influenced. In order to investigate the exact orbital effect of $V^{3+}$-ions, it is essential to substitute $V^{3+}$-ions by an ion without magnetism or with different magnetic/orbital properties gradually. $Mn^{3+}$ ($3d^4$, S=2) ion has a different orbital effect on $e_g$ orbitals, ($3z^2$-$r^2$), which is extended in the basal $ab$-plane in an octahedral environment as the tetragonal phase (c > $a$) in $Mn_3O_4$ [21-23]. Hence, doping $Mn^{3+}$ ions on the $V^{3+}$-site provided a perfect example to study the orbital properties of $V^{3+}$-ions . In this work, we modulated constituent of B-site ions and studied the structural and magnetic properties of $Mn_{1+x}V_{2-x}O_4$ (0 < $x$ ≤ 1) by magnetization, heat capability, XRD and neutron diffraction measurements, and the first principle simulation. The results confirmed the phase splitting between the crystal structure and magnetic structure. The magnetic transition still exists while crystal structure maintains tetragonal phase ($c > a$) in the whole temperature range for high doping ($x \geq 0.4$). The transition temperature decreases with increasing $x$, which implies the competition between exchange interactions $J_{AB}$ and $J_{BB}$ becomes stronger. Two other noncollinear phase transitions related to $V^{3+}$ and $Mn^{3+}$ ions are also detected.

## 2. Sample synthesis and experiments

Polycrystalline samples of $Mn_{1+x}V_{2-x}O_4$ (0 < $x$ ≤1) were synthesized by solid state reaction. Stoichiometric mixtures of MnO, $V_2O_3$, and $V_2O_5$ were ground together and calcined under flowing Ar in sealed quartz tube at 900°C for 100h. The magnetization was measured by a Quantum Design Physical Property Measurements System (PPMS) with an applied field H=100 Oe. The specific heat measurements were also performed on PPMS. The XRD patterns were recorded by a HUBER imaging-plate Guinier camera 670 with Ge monochromatized Cu $K_{\alpha1}$ radiation. Data were collected at temperatures from 10 K to 300 K with a cryogenic helium compressor unit. Neutron powder diffraction (NPD) measurements were performed at the POWGEN [24] diffractometer at the Spallation Neutron Source (SNS), and the neutron powder diffractometer (HB-2A) at the High Flux Isotope Reactor (HFIR), Oak Ridge National Laboratory (ORNL). ~1.5 g powder of $Mn_{1+x}V_{2-x}O_4$ ($x$ = 0.1, 0.2, and 0.3) were loaded in V cans in the sample changer at POWGEN, and the patterns were collected from 10 K to 90 K. ~ 3 g powder of $Mn_{1+x}V_{2-x}O_4$ ($x$ = 0.1, 0.2, and 0.3) were measured at HB-2A at 1.5 K and 150 K with both wavelengths of 1.54 Å and 2.41 Å, and the collimation was set as open-21'-12'. The shorter wavelength was used to investigate the crystal structures, while the longer wavelength provided lower Q coverage with better resolution that was important for investigating the magnetic structures of the material.

The diffraction data were analyzed by the refinement program FullProf.

## 3. Results

### 3.1) XRD and NPD

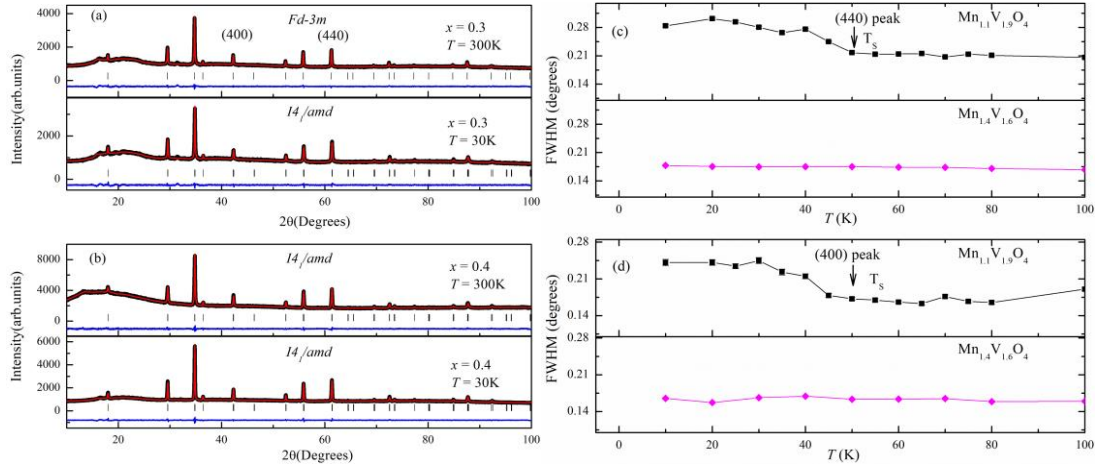

FIG.1. (Color online) XRD data of $Mn_{1+x}V_{2-x}O_4$ ($0.1 \leq x \leq 1$). Rietveld refinements of $Mn_{1.3}V_{1.7}O_4$ (a) and $Mn_{1.4}V_{1.6}O_4$ (b) by FULLPROF at 300 K and 30 K, respectively. The crystal structures of $Mn_{1.3}V_{1.7}O_4$ are cubic at 300K and tetragonal at 30K, while the lattice of $Mn_{1.4}V_{1.6}O_4$ keeps as tetragon at both temperatures. The temperature dependence of FWHM of (400) (c) and (440) diffractions (d) for $Mn_{1.1}V_{1.9}O_4$ and $Mn_{1.4}V_{1.6}O_4$, respectively.

The composition- and temperature-dependence of XRD powder diffraction were measured from 10 K to 300 K and analyzed by the Rietveld refinement with the FULLPROF program [25]. Figures 1(a) and 1(b) present the data and refinement of $Mn_{1+x}V_{2-x}O_4$ ($x$ = 0.3 and 0.4) at 30 K and 300 K, respectively, and the refinement parameters of space group, lattice constants, and atomic positions are included in Table I. As $MnV_2O_4$, the structure transition from cubic $Fd$-$3m$ to tetragonal $I4_1/amd$ symmetry was observed for $Mn_{1.3}V_{1.7}O_4$, however, the lattice constant ratios are different: $c/a < 1$ for $MnV_2O_4$ and $c/a > 1$ for $Mn_{1.3}V_{1.7}O_4$. Actually, It is worth mentioned that $Mn_3O_4$ has also the similar tetragonal phase with $c/a > 1$. Based on the results of our refinement, the change of lattice constant ratios from $c/a < 1$ to $c/a > 1$ should be attributed to the orbital order competition between $Mn^{3+}$- and $V^{3+}$-ions [26, 27]. On the other hand, the $x > 0.3$ system keeps the tetragonal ($c/a > 1$) structure from 300 K to 10 K.

Table 1. Structural parameters for the $x$ = 0.3 at 300 K ($Fd$-$3m$) and 30 K ($I4_1/amd$), and $x$=0.4 at 300 K ($I4_1/amd$) and 30 K ($I4_1/amd$), respectively.

|  | Atom | Site | x | y | Z |
|---|---|---|---|---|---|
| **$x$ = 0.3, $T$ = 300K** | | | | | |
| *Fd-3m* | Mn (1) | 8a | 0.125 | 0.125 | 0.125 |
| $a$= 8.551 Å | V | 16d | 0.5 | 0.5 | 0.5 |
| $b$= 8.551 Å | Mn (2) | 16d | 0.5 | 0.5 | 0.5 |
| $c$= 8.551 Å | O | 32e | 0.740 | 0.740 | 0.740 |
| $\chi^2$= 0.243 | | | | | |
| **$x$ = 0.3, $T$ = 30K** | | | | | |
| *I4$_1$/amd* | Mn (1) | 4a | 0.5 | 0.25 | 0.125 |
| $a$= 6.033 Å | V | 8c | 0.25 | 0.75 | 0.25 |
| $b$= 6.033 Å | Mn (2) | 8c | 0.25 | 0.75 | 0.25 |
| $c$= 8.549 Å | O | 32e | 0.5 | 0.984 | 0.267 |
| $\chi^2$= 0.256 | | | | | |
| **$x$ = 0.4, $T$ = 300K** | | | | | |
| *I4$_1$/amd* | Mn (1) | 4a | 0.5 | 0.25 | 0.125 |
| $a$= 6.035 Å | V | 8c | 0.25 | 0.75 | 0.25 |
| $b$= 6.035 Å | Mn (2) | 8c | 0.25 | 0.75 | 0.25 |
| $c$= 8.549 Å | O | 32e | 0.5 | 0.992 | 0.277 |
| $\chi^2$= 0.592 | | | | | |
| **$x$ = 0.4, $T$ = 30K** | | | | | |
| *I4$_1$/amd* | Mn (1) | 4a | 0.5 | 0.25 | 0.125 |
| $a$= 6.033 Å | V | 8c | 0.25 | 0.75 | 0.25 |
| $b$= 6.033 Å | Mn (2) | 8c | 0.25 | 0.75 | 0.25 |
| $c$= 8.545 Å | O | 32e | 0.5 | 0.984 | 0.269 |
| $\chi^2$= 0.336 | | | | | |

Moreover, the temperature-dependence of (440) and (400) reflections are investigated in Figs. 1(c) and 1(d), respectively. The full width at half maximum (FWHM) of both peaks increases at $T_S$ ~ 54 K for $Mn_{1.1}V_{1.9}O_4$, which suggests a structural transition and is consistent with the heat capacity measurement. Meanwhile, the FWHM of $Mn_{1.4}V_{1.6}O_4$ remains constant in the entire temperature range, indicating no structural transition, Fig. 1(b) and Table 1.

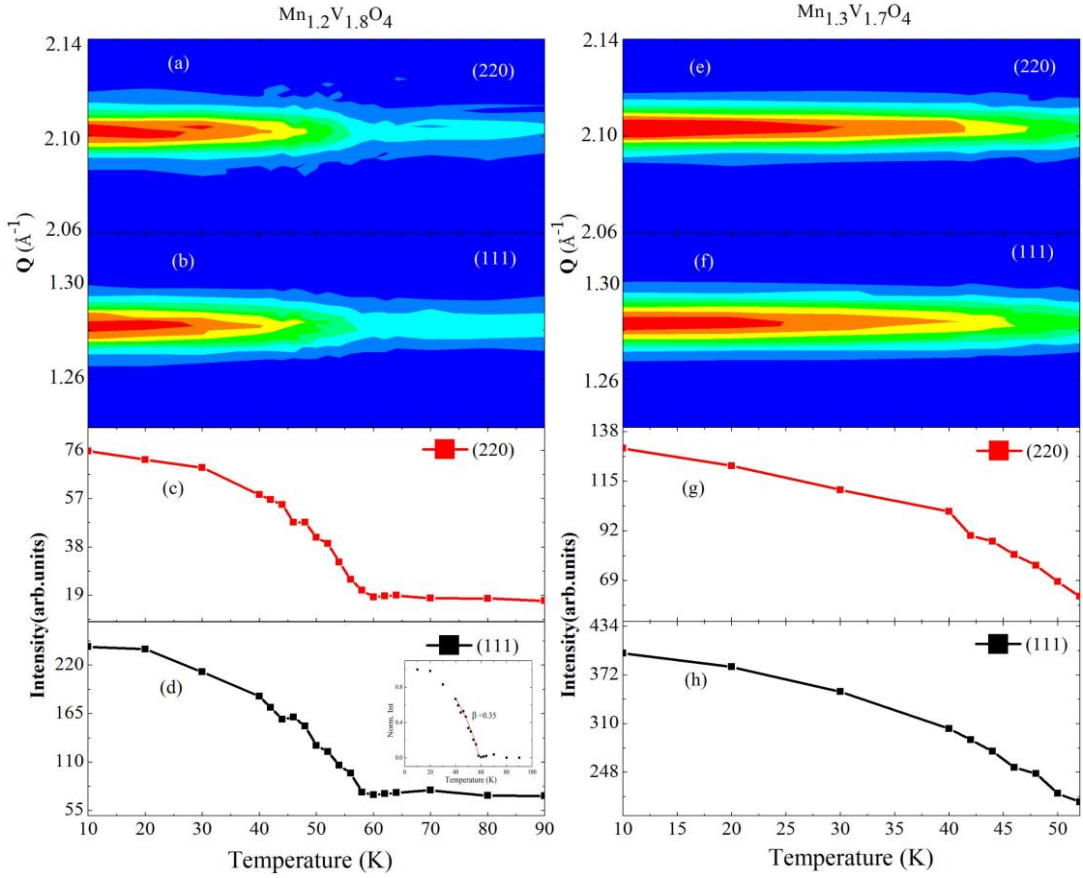

FIG.2. (Color online) The temperature-dependence of the (220) and (111) Bragg peaks of $Mn_{1+x}V_{2-x}O_4$ for $x = 0.2$ [(a), (b)] and $x = 0.3$ [(e), (f)] by the neutron powder diffractometer POWGEN, respectively. (c)-(d), (g)-(h) are the related (220) and (111) peak intensities for $x = 0.2$ [(c), (d)] and $x = 0.3$ [(g), (h)]. Inset: The power law fits of (111) for $x = 0.2$.

The NPD is also performed on POWGEN and HB-2A, ORNL, which will be helpful to further understand the lattice and magnetic structure. Figure 2 demonstrated the (220) and (111) reflections of $Mn_{1.2}V_{1.8}O_4$ at different temperatures. These two Bragg peaks include both structural and magnetic intensities. As shown in Fig. 2, the peak broadening indicates a lattice distortion upon cooling down, and the obvious increase on peak intensities at about 60K suggests the occurrence of magnetic order. As described in $MnV_2O_4$ [19] and $Mn_{1+x}Co_xV_2O_4$ [10], both a magnetic transition from PM to CF and a structural transition from cubic to tetragonal lattice should happen at the same temperature. In order to examine the phase transitions more clearly, the temperature dependence of intensities are plotted in Figs. 2(c)-(d), (g)-(h). For the critical behavior of the magnetic sublattice for $Mn_{1+x}V_{2-x}O_4$, the temperature dependence of the magnetic peak is analyzed, Fig. 3(f) (inset). A power law (Eq. (1)) is applied to fit the integrated intensities $I$ of the magnetic diffraction peaks near the transition temperature,

$$I = I_0 \left(1 - \frac{T}{T_N}\right)^{2\beta} \tag{1}$$

where $T_N$ is the Néel temperature, $I_0$ is the integrated intensity at base temperature, and $\beta$ is the order parameter critical exponent. The obtained $\beta$ is 0.35 for $Mn_{1.2}V_{1.8}O_4$, which is close to $\beta_2$ (0.34) of $MnV_2O_4$ [19] and the 3D Heisenberg ($\beta = 0.36$) model.

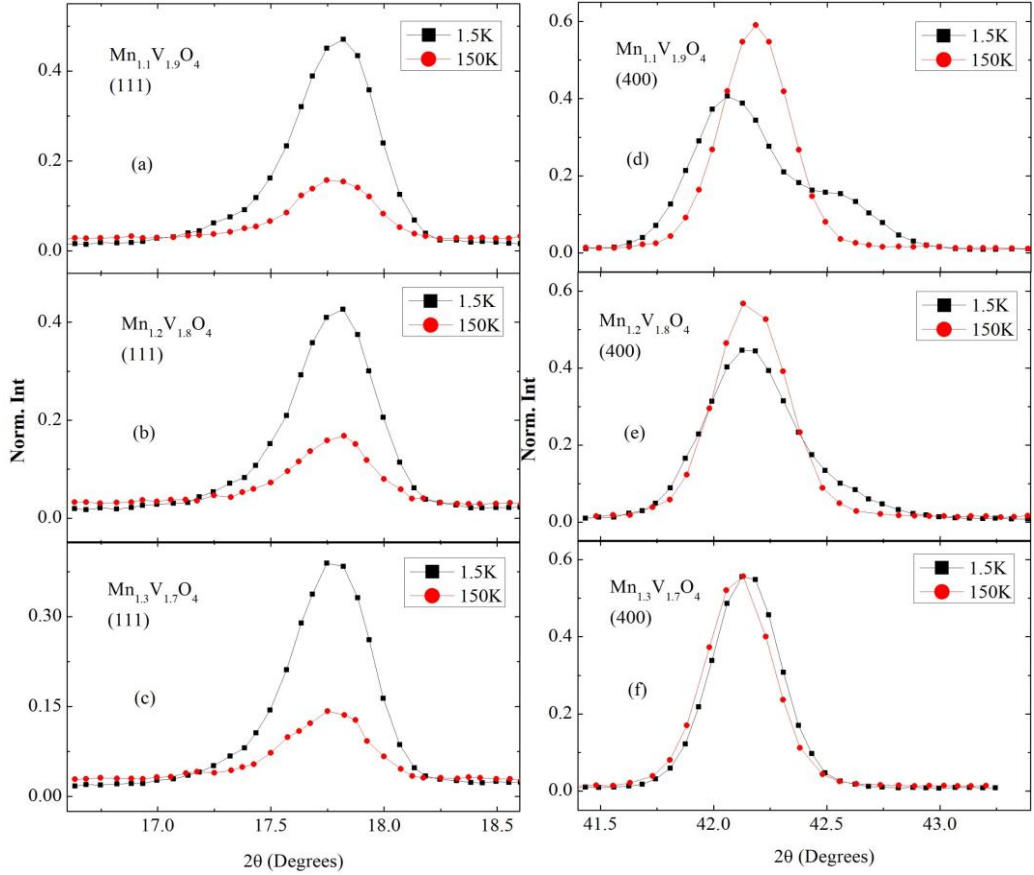

FIG.3. (Color online) (a)-(f), the (111) and (400) Bragg diffraction of $x$ = 0.1, 0.2, 0.3 at 1.5 K and 150 K by the neutron powder diffractometer, HB-2A, $\lambda$ = 1.54Å. The diffraction splitting from cubic to tetragonal phase was clearly demonstrated.

As shown in Fig. 3, the high resolution NPD measurements on $Mn_{1+x}V_{2-x}O_4$ at 1.5 K and 150 K with $\lambda$ = 1.54Å were performed. The intensity of (111) magnetic peak decreases clearly for all samples, coincided with the formation of magnetic order. The (400) Bragg diffraction of $Mn_{1.1}V_{1.9}O_4$ at 150K with cubic phase clearly split into (004) and (220) Bragg peaks at 1.5K, which recommended a structural phase transition from cubic phase to tetragonal phase. Comparing with $Mn_{1.2}V_{1.8}O_4$ and $Mn_{1.3}V_{1.7}O_4$ (Figs. 3(e) and 3(f)), one should notice the degree of crystal distortion decreased as $x$ increase.

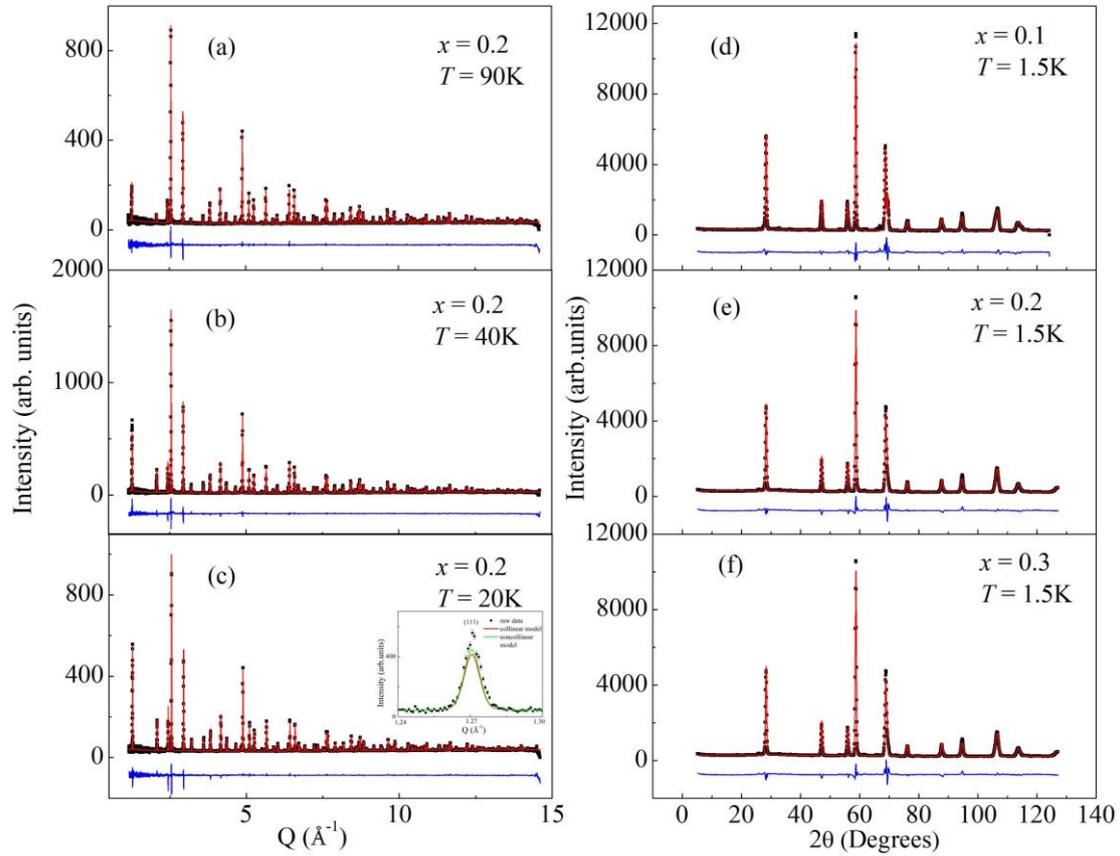

FIG.4. (Color online) (a)-(c) Temperature-dependence of NPD data for $Mn_{1.2}V_{1.8}O_4$ by POWGEN, and (d)-(f) composition-dependence of NPD data for $Mn_{1+x}V_{2-x}O_4$ ($x$ = 0.1, 0.2 and 0.3) with $\lambda$=2.41Å at 1.5K by HB-2A, respectively. Black dots are the raw data, red lines are Rietveld refinements and blue lines are the differences between observed and calculated intensities. Inset: fitting results of (111) peak by different models.

In Figs. 4(a)-4(c), $Mn_{1.2}V_{1.8}O_4$ exhibits a cubic structural phase with PM at 90K, which is compatible with the XRD results. Then it undergoes the first phase transition with the decreasing temperature, and a tetragonal ($c > a$) structure with spins of $Mn^{2+}$ and $V^{3+}$ aligning antiparallel to each other was obtained at 40K. Such a model is good to fit our NPD data. With further cooling temperature, a model of spins of $V^{3+}$ canted from c axis and that of $Mn^{3+}$ kept disorder was used to refine 20K data, which is good agreement with our experimental data. However, the indicator of the CF-to-NCF transition, the magnetic diffraction of (002), is not captured due to the weak intensity of the powder average effect and high absorption effect of $V^{3+}$ ions [20]. Hence, the models of the $V^{3+}$ spins antiparallel to $Mn^{2+}$ spins and the $V^{3+}$ spins canted in *ab*-plane were applied. The comparison of the canted and non-canted $V^{3+}$-ions model on (111) diffractions is presented in inset of Fig. 4(c), which clearly demonstrates that the canted $V^{3+}$-ions model agreed with the measurement better. To identify the change of canting angle of $V^{3+}$ spins with *x* increasing, the high resolution neutron powder diffraction measurements were performed on $Mn_{1+x}V_{2-x}O_4$ (*x* = 0.1, 0.2 and 0.3) at 1.5K, and the models of the collinear/noncollinear $Mn^{3+}$ spins were tested. For the collinear model, the NPD data couldn't be fitted very well with a ferromagnetism of $Mn^{2+}$- $Mn^{3+}$ and a large $Mn^{3+}$ moment; while the model of noncollinear $Mn^{3+}$ spins was not suitable, either. So the model of spins of $V^{3+}$ canted from c axis and that of $Mn^{3+}$ kept disorder was used to refine HB-2A data. The refinement of Figs 3(d)-3(f) demonstrated that the canting angles decreased with increasing *x*, (*x* = 0.1, 43.47°; *x* = 0.2, 35.75°; *x* = 0.2, 32.21°), respectively, which agrees with $Mn_{1+x}Co_xV_2O_4$ very well [10] and a smaller $V^{3+}$-$V^{3+}$ bond length was recommended to $MnV_2O_4$.

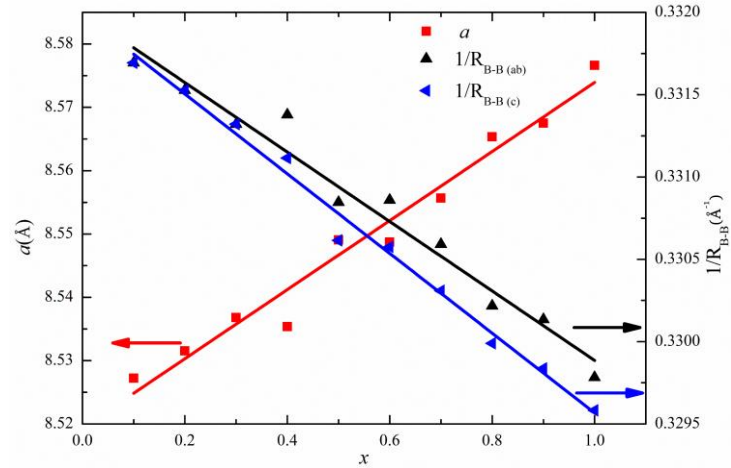

FIG.5. (Color online) The composition-dependence of the lattice parameter *a* and the related $1/R_{B-B}$ along different directions for $Mn_{1+x}V_{2-x}O_4$ (0.1 ≤ *x* ≤ 1) at room temperature.

Previous researches have confirmed that both $MnV_2O_4$ and $Mn_3O_4$ were normal spinel structure with $Mn^{2+}$ ions at A-sites and $V^{3+}$-/$Mn^{3+}$-ions at B-sites [28, 29], while the $Fe_2VO_4$, $Co_2VO_4$ and $Mg_2VO_4$ were inverse spinel [30-32]. Since Mn-doping affects both orbital effects and valences of Mn and V ions in $Mn_{1+x}V_{2-x}O_4$, both normal spinel and inverse-spinel lattices were applied to refine the XRD and neutron powder diffractions. Although the fitting of $Mn_{1+x}V_{2-x}O_4$ kept as spinel lattice up to *x* = 0.9 for XRD results, there is no big difference of the fitting parameter, $\chi^2$, for $Mn_2VO_4$.

Therefore, the first-principle theory was performed on $Mn_2VO_4$ to analyze the accurate distribution of ions with the tetragonal lattice and would be discussed later. From the first-principle calculation, the average atomic potential energy of inverse spinel is -7.98939eV, while the normal spinel had the energy of -8.00222eV with 12.8meV lower. Therefore, $Mn_2VO_4$ was identified as a normal spinel with the tetragonal lattice.

In $Mn_{1-x}Co_xV_2O_4$ [20, 33], the smaller $Co^{2+}$ ions not only increase chemical pressure in the system, but also shorten the Co-Co, V-V and Co-V bond lengths. As $R_{V-V}$ is an indicator of weakening the orbital effect of $V^{3+}$-ions at B-site, the composition-dependence of the lattice constant and $1/R_{B-B}$ are shown to determine the orbital effects of $V^{3+}$-ions and $Mn^{3+}$-ions with $Mn^{3+}$-doping, Fig. 5. The parameters for $x \geq 0.4$ are multiplied tetragonal $a$ by $\sqrt{2}$, and both the lattice parameter and B-B bond length increase linearly with $x$. The increasing distance will reduce the chemical pressure and structural isotropy, which is just contrary to $Mn_{1-x}Co_xV_2O_4$. Thus, the doped $Mn^{3+}$ ions localize the electrons and weaken the superexchanges of $J_{AB}$ and $J_{BB}$ to reduce $T_1$, meanwhile, the disappearance of structural phase transition leads to the significant drop on $T_1$ between $x = 0.4$ and 0.5.

3.2) Magnetization and heat capacity

The temperature-dependent ZFC (zero-field-cooled) and FC (field-cooled) magnetizations of $Mn_{1+x}V_{2-x}O_4$ ($0.1 \leq x \leq 1$) under 100 Oe magnetic field are shown in Fig. 6. As temperature decreases, a clear rise is observed at $T_1$, which indicates the magnetic phase transition from PM to CF order [18, 28], We also measured the temperature dependence of the specific heat ($C_P$–T curve) in zero magnetic field for all the samples, and one peak was observed in each $C_P$–T curve at the same temperature of $T_1$. And the peak shape broadens with the $Mn^{3+}$ content increasing. When $x \leq 0.3$, we can observe a sharp peak, which corresponds to the lattice and magnetic transition simultaneously occurring at $T_1$. When $0.4 \leq x \leq 1$, the broad peak only corresponds to the magnetic order transition. More anomaly could be observed at lower temperature in the high doping ($x \geq 0.4$) ZFC curve. In order to probe the magnetic phase transition of $Mn_{1+x}V_{2-x}O_4$ ($0.1 \leq x \leq 1$) in more detail, the derivative of the ZFC magnetization is shown in Fig. 7. Besides $T_1$, another peak marked as $T_2$ was observed in each curve below the $T_1$. $T_2$ should correspond to the spin canting transition of $V^{3+}$ as $MnV_2O_4$ [19]. Both $T_1$ and $T_2$ decreased with the $Mn^{3+}$ content increasing, and $T_2$ decreased fast at $x = 0.2$. Since the magnetic transition from CF-to-NCF is related to the orbital ordering of $V^{3+}$ ions in $MnV_2O_4$ [19], the sharp decreasing $T_2$ suggests the dilute effect from the $Mn^{3+}$-ions. For $x \geq 0.5$, the third peak was observed, which may be induced by the collinear to noncollinear transition of $Mn^{3+}$. Because in $Mn_3O_4$, the ground state of $Mn^{3+}$ spin is noncolinear without the intermediated collinear ferrimagnetism [21, 25, 26, 33]. Both the noncollinear phase transition of $T_2$ and $T_3$ were associated with competition between $J_{AB}$ and $J_{BB}$ and spin-orbit coupling of magnetic $Mn^{3+}$- and $V^{3+}$-ions at B sites. We would discuss the mechanism for these transitions concretely later.

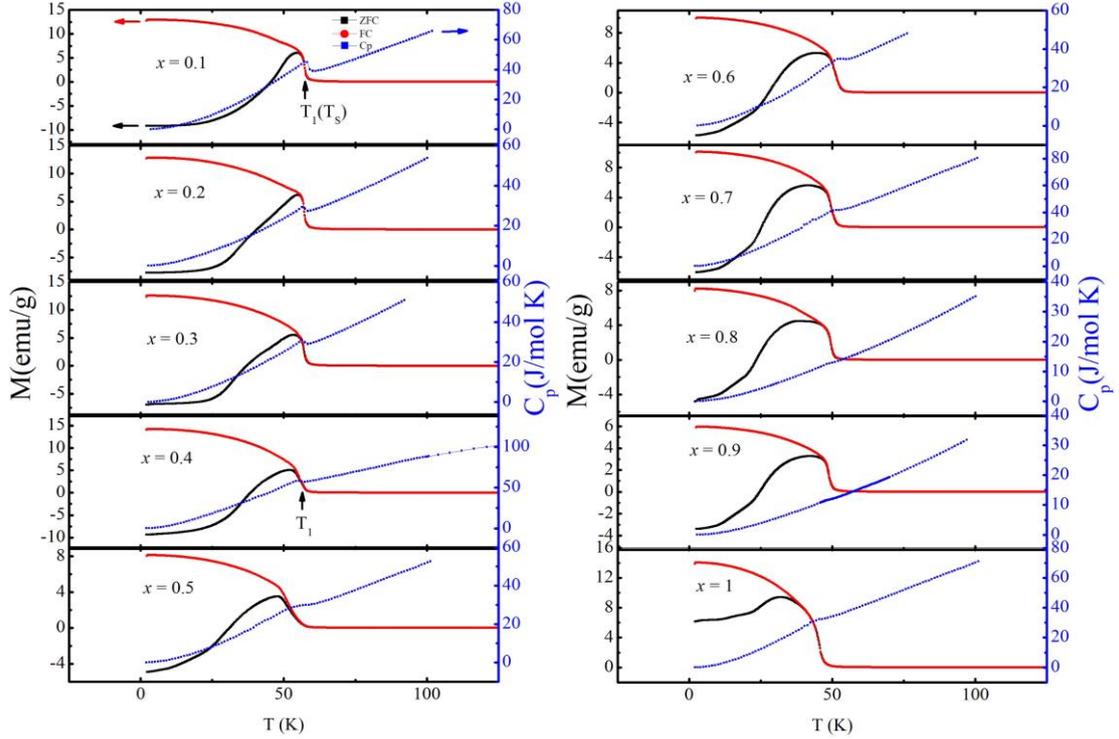

FIG.6. (Color online) The temperature dependence of magnetization and specific heat for $Mn_{1+x}V_{2-x}O_4$ ($0.1 \leq x \leq 1$). Black and red lines are the results of the zero-field-cooled (ZFC) and field-cooled (FC) measurements, respectively. The blue line represents the specific heat data. For $x \leq 0.3$, $T_1$, which corresponds to the PM to CF transition temperature, is equal to the structural transition temperature $T_S$ (Cubic to tetragonal). However, for $x \geq 0.4$, $T_1$ remains while $T_S$ disappears.

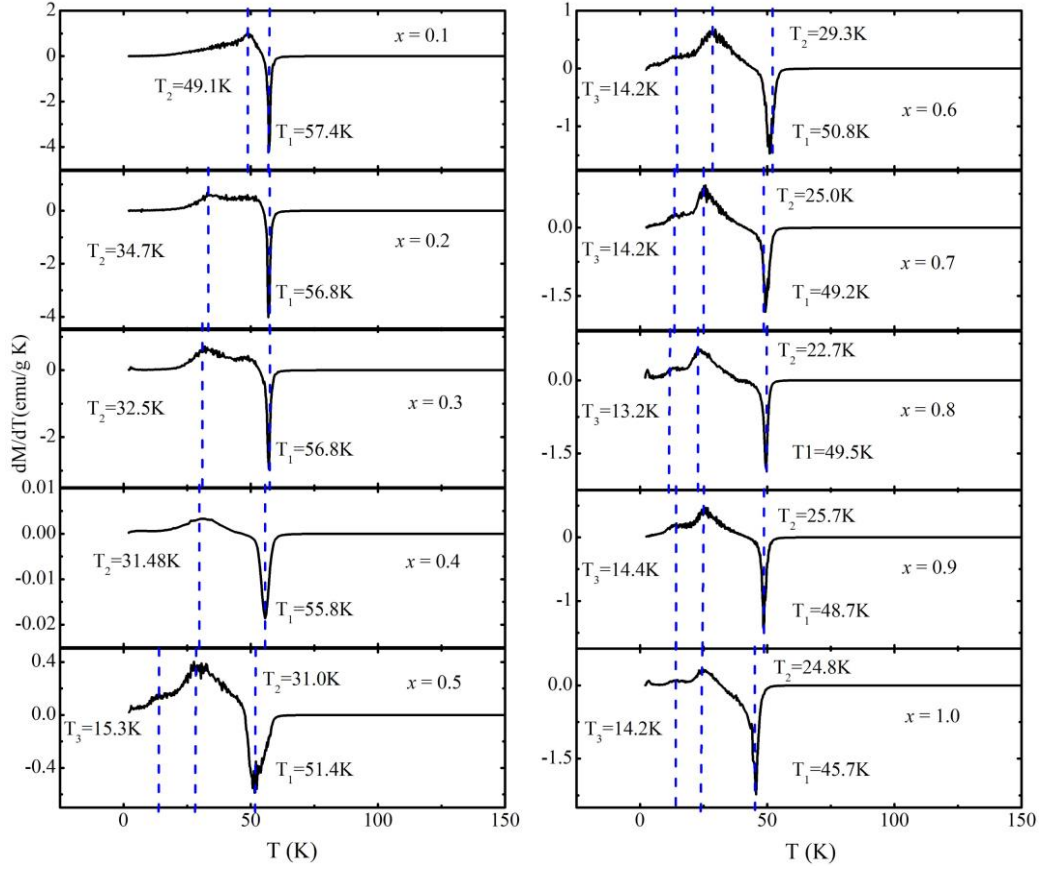

FIG.7. (Color online) The first derivative ZFC curve of $Mn_{1+x}V_{2-x}O_4$ ($0.1 \leq x \leq 1$). $T_1$ represents the transition temperature of PM to CF, $T_2$ corresponds to the transition temperature from CF to the first noncollinear ferrimagnetism (NCF1), $T_3$ is the temperature from the first non-collinear ferrimagnetism to the second noncollinear ferrimagnetism (NCF2).

3.3) First principle calculation

In order to study the orbital effects on the related superexchange, partial density-of-states (DOS) of $Mn_2VO_4$ were calculated with the normal spinel and inverse-spinel lattice, respectively, Fig. 8. First-principle calculations of the electronic structures for normal spinel and inverse-spinel $Mn_2VO_4$ were carried out with the Vienna ab initio simulation package (VASP) [34]. The projector augmented-wave method within the density-functional theory was used to describe the interactions between ions and electrons [35], and the potential energy can be determined without using any empirical input. Specifically, PBE [36] functional under the general gradient approximation and a plane-wave cutoff of 400 eV were used. To accurately describe the atomic structure of $AB_2O_4$ with chemical occupation disorder, a super cell containing 448 atoms (64 $A^{2+}$, 128 $B^{3+}$, 256 $O^{2-}$) was adopted for both the normal spinel and inverse-spinel lattice, with the occupation disorders being imitated via random number generation. For the spinel lattice, the 64 $A^{2+}$ sites contain only $Mn^{2+}$ ions, whereas the 128 $B^{3+}$ sites are randomly filled with 64 $Mn^{3+}$ and 64 $V^{3+}$ ions. For the inverse-spinel lattice, the 64 $A^{2+}$ sites are randomly filled with 32 $Mn^{2+}$ and 32 $V^{2+}$ ions, whereas the 128 $B^{3+}$ sites are randomly filled with 96 $Mn^{3+}$ and 32 $V^{3+}$ ions. For accurate electronic structure calculations, the Brillouin zone was sampled with a 3 × 3 × 3 k-point mesh. For comparison, the partial electronic density of states for different elements were normalized by the number of atoms for each element. Although the calculation was based on $Mn_2VO_4$, the DOSs of $Mn^{3+}$-/$V^{3+}$-ions in normal spinel and inverse spinel lattice were similar in the whole system: i) The *s*- and *p*-orbital DOSs were similar for both normal spinel and inverse-spinel; ii) difference between the normal spinel and inverse-spinel. For normal spinel structure (I), the $e_g(3z^2-r^2)$ DOS of $Mn^{3+}$-ion had a sharp peak around fermi surface, while $t_{2g}$ orbitals were dominant in inverse spinel (II). Unlike inverse spinel, the $t_{2g}$ DOS of normal spinel $V^{3+}$-ion was larger than $e_g$ orbitals. As expected, the result of normal spinel structure was consistent with the orbital ordering in previous study [17, 23].

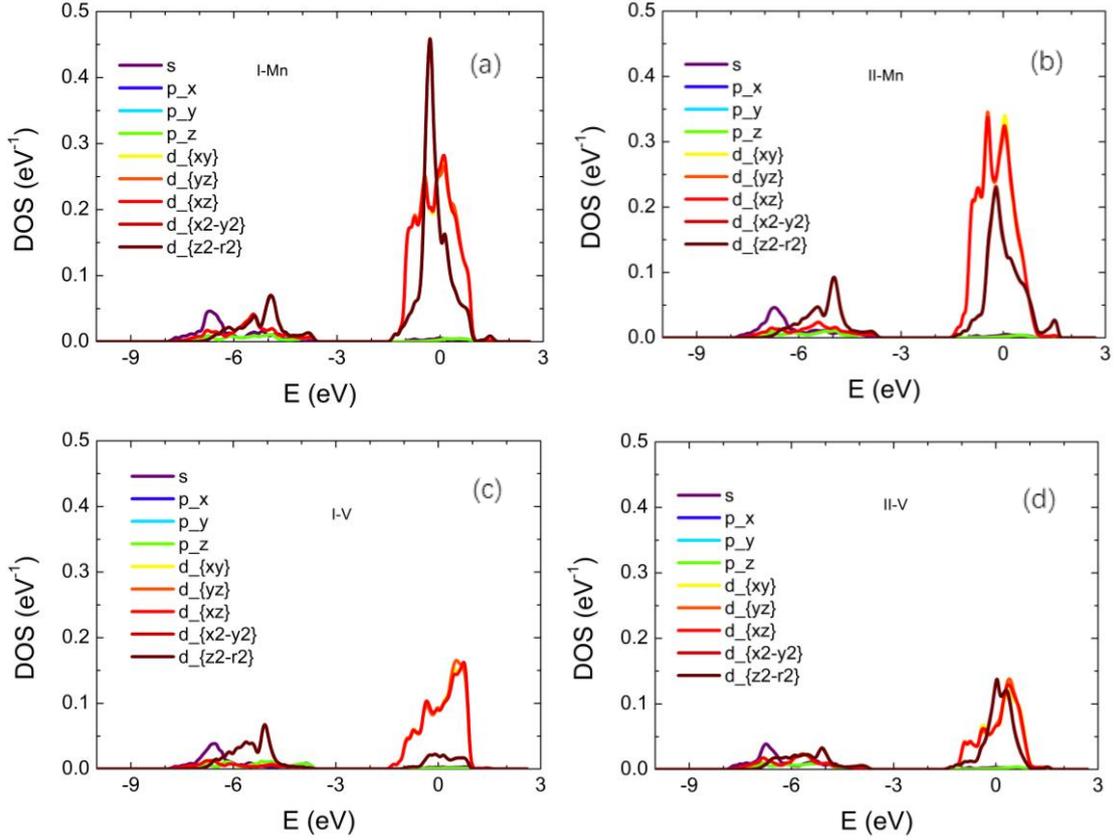

FIG.8. (Color online) Density-of-states (DOS) of $Mn_2VO_4$ from DFT calculations. (a) and (b), DOS of Mn calculated by the normal spinel model (I) and inverse spinel model (II), respectively. (c) and (d), DOS of V calculated by the normal spinel model (I) and inverse spinel model (II), respectively.

3.4) Phase diagram

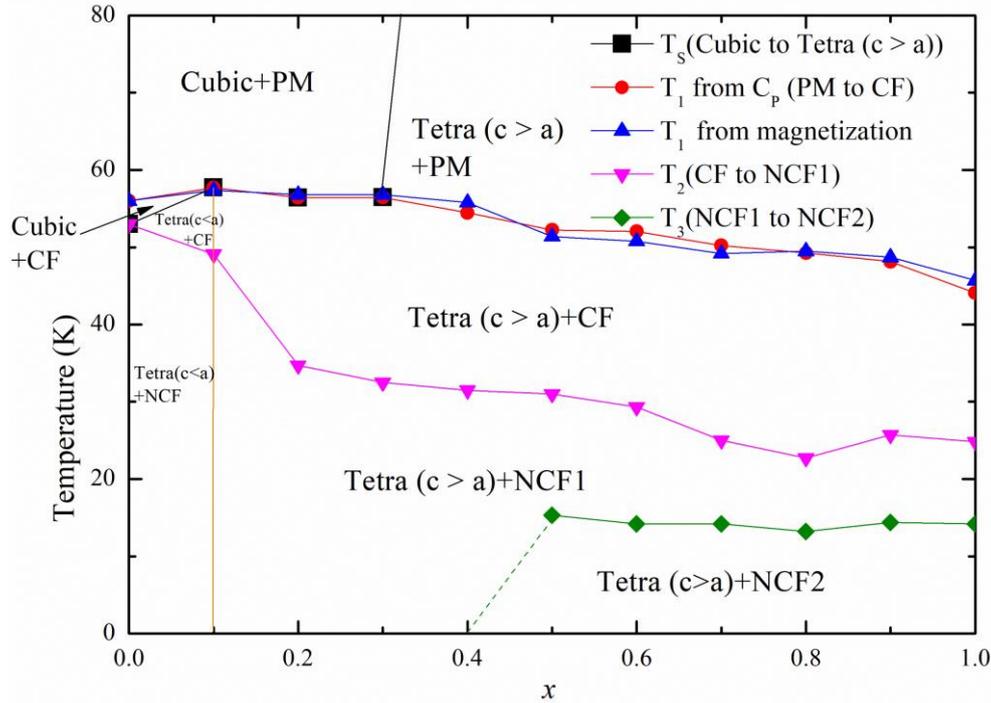

FIG.9. (Color online) The temperature versus $Mn^{3+}$ content ($x$) phase diagram of $Mn_{1+x}V_{2-x}O_4$. $T_S$ is the cubic-to-tetragonal lattice transition temperature (black lines and squares) obtained from XRD and heat capacity measurements. $T_1$ is the paramagnetic-to-collinear ferrimagnetic phase transition temperature (red line/dots and bule line/triangles) obtained from heat capacity and magnetization measurements. $T_2$ is the CF-NCF1 phase transition temperature (pink line/triangles) from magnetization measurement, where the spins of $V^{3+}$ ions cant in $ab$-plane with that of $Mn^{3+}$ ions align along c axis. $T_3$ is the temperature from NCF1 to NCF2 (olive line/rhombuses), and the spins of $Mn^{3+}$ ions cant in $ab$-plane. The $MnV_2O_4$ data are from reference [29].

Combining the heat capacity, magnetization, XRD and neutron diffraction data, the temperature(T) vs. composition($x$) phase diagram of $Mn_{1+x}V_{2-x}O_4$ was constructed in Fig. 9. This complicated phase diagram clearly presents $Mn^{3+}$-doping effects on the Jahn-Teller distortion, the spin-orbit coupling, the competition of exchange interactions among different sites and chemical pressure in the system. $T_S$ corresponds to the structural phase transition from cubic to tetragonal phase, which was obtained from heat capacity and XRD. $T_1$ represents the magnetic phase transition temperature from PM to CF. $T_2$ and $T_3$ are the CF-NCF1 and NCF1-NCF2 phase transition temperatures, where the spins of $V^{3+}$ and $Mn^{3+}$ cant from *c*-axis in the *ab*-plane, respectively.

## 4. Discussions

Although the same tetragonal phase, *I4$_1$/amd*, is observed in the whole $Mn^{3+}$-doped $MnV_2O_4$ system, the low doping compounds ($0 \leq x \leq 0.1$) have $c/a < 1$ as $MnV_2O_4$ and the high doping compounds ($x > 0.1$) have $c/a > 1$ as $Mn_3O_4$. Moreover, $T_S$ is not observed up to 300K above $x = 0.3$ and should be determined by the collaboration of the orbital effects of $Mn^{3+}$- and $V^{3+}$-ions.

The structural phase transition in $MnV_2O_4$ can be associated with the orbital ordering of $V^{3+}$ ions on the B-sites [18, 19, 37, 38]. Since the *d* orbitals of $V^{3+}$ ions split into three low-energy $t_{2g}$ orbitals and two high-energy $e_g$ orbitals, two outer shell electrons of $V^{3+}$ ion localize on the threefold degenerate $t_{2g}$: One electron occupies the *xy* orbital, and the other has the freedom of filling the *yz* or *zx* orbital or both of them with partial probabilities along *c*-axis [17], which agrees with the dominant contribution from the $t_{2g}$ orbitals of $V^{3+}$ DOS, Fig. 8(c). Hence, the shielding effect of electrons on the central ion is weak along *c*-axis, and the central nucleus can suffer an attractive force from the *c*-direction, which distorts the $V^{3+}$-lattice along *c*-axis. The Jahn-Teller effect not only confines at the V-pyrochlore, but also interferes the whole crystal to drive the structural transition from cubic to tetragonal (c < a) phase.

On the other hand, the $Mn^{3+}$ ($3d^4$) ions locate at center of a tetragonal $MnO_4$ with high spin configuration and the only electron occupies the $e_g$ $3z^2$-$r^2$ orbit. Just as the dominated $Mn^{3+}$ DOS of $d(3z^2$-$r^2)$ around Fermi surface, Fig. 8(a). The similar tetragonal phase (c > a) is observed in $Mn_3O_4$, and the transition of cubic to tetragonal occurs at 1443 K [39]. Therefore, the structural phase of $Mn_{1+x}V_{2-x}O_4$ for low $Mn^{3+}$ doping ($x \leq 0.3$) at $T_S$ can be regarded as the result of orbital ordering competition between $Mn^{3+}$ and $V^{3+}$ with the compression and elongation of $BO_6$ octahedron respectively. Comparing with the diagonal orbitals of $V^{3+}$-ions, $Mn^{3+}$-ions occupy the orbital along axis and more easily influence the crystal distortion. As a consequence, a c > a tetragonal phase appears at $T_S$. For high $Mn^{3+}$ doping ($x \geq 0.4$), the lattice keeps a tetragonal phase up to room temperature, presenting that the orbital ordering of $Mn^{3+}$-ions suppressed $V^{3+}$ orbital completely.

Furthermore, the magnetic phase transition temperatures are strongly related to the Hamiltonian of exchange energy. The full magnetic Hamiltonian may include the isotropic Heisenberg exchange constants $J_{ij}$ and the single ion anisotropies $D_i$, where d is the axis of the anisotropy so that the spin Hamiltonian can be written as follows:

$$\begin{aligned}H = &J_{AB}\sum_{(p,q)(i,j,k,l)}(S_p+S_q)\cdot(S_i+S_j+S_k+S_l)\\&+J_{BB}^{ab}(\sum_{i,j}S_i\cdot S_j+\sum_{k,l}S_k\cdot S_l)+J_{BB}^{c}\sum_{(i,j)(k,l)}(S_i+S_j)\cdot(S_k+S_l),\\&+D_A\sum_{r=p,q}(\hat{z}\cdot S_r)^2+D_B\sum_{s=i,j,k,l}(\hat{u}_s\cdot S_s)^2\end{aligned} \quad (2)$$

where $J_{AB}$, $J_{BB}^{ab}$, and $J_{BB}^{c}$ are nearest-neighbor interactions. The inequivalent A-sites are given by subscripts p and q, and the inequivalent B-sites are given by subscripts i, j, k and l. For the A-site spins, the easy-axis anisotropy, $D_A$, is along the c-axis while for the B-site spins, the easy-axis anisotropy, $D_B$, is along the local <111> direction.

At $T_1$, the spins of $Mn^{2+}$ and $V^{3+}$ align antiparallel to each other and form a collinear magnetic ordering phase, while the spins of $Mn^{3+}$ at B-sites can still keep disordered. Since the antiferromagnetic exchange energy of $Mn^{3+}$-$Mn^{3+}$ ($J_{BB}^{ab}$) is much stronger than the $Mn^{3+}$-$Mn^{2+}$ coupling ($J_{AB}$), a collinear magnetic structure can't form under frustration. There are two paths to achieve magnetic interactions between nearest-neighbor $Mn^{3+}$ ions [40, 41]: i) the direct exchange interactions between the neighboring $t_{2g}$ orbitals which are strongly antiferromagnetic; ii) the superexchange interaction involving neighboring $e_g$ electrons and middle oxygen's $2p$ orbital, which is weakly ferromagnetic. Therefore, the exchange interaction of $Mn^{3+}$-$Mn^{3+}$ exhibits strongly antiferromagnetic in $ab$-plane, and $J_{BB}^{c}$ is weak along the $c$-axis due to elongation of $BO_6$ octahedra. With the mechanism such as interplay of $J_{AB}$ and $J_{BB}$, a collinear Néel phase can not appear in $Mn_3O_4$, while a noncollinear ferrimagnetic YK phase presents instead [21, 41]. However, $J_{BB}$ is not strong compared with $J_{AB}$ in $MnV_2O_4$ and spins of $V^{3+}$ ions can be aligned parallel to each other to produce collinear long-range ordering, so the spins of $Mn^{3+}$ remain disordered but $V^{3+}$ ordered at $T_1$ in $Mn_{1+x}V_{2-x}O_4$. With more $Mn^{3+}$ doping, the total exchange energy will decrease by reducing the first part of Hamiltonian mainly. Thus, the long-range spin order can be disturbed by thermal perturbation easily and $T_1$ decreased with $x$ increasing.

$T_2$ and $T_3$ in Fig. 9 are derived from the canting of the spins on B sites, $Mn^{3+}$ ions

and $V^{3+}$ ions, respectively. The first NCF phase (NCF1) is from the canting spin of $V^{3+}$ and the second NCF phase (NCF2) is from that of $Mn^{3+}$. The similar noncollinear magnetic phase transitions are both found in $MnV_2O_4$ [19] and $Mn_3O_4$ [42]. In the NCF phase, the magnetic unit cell possesses two $Mn^{2+}$ spins aligned along $c$-axis, while the tetrahedra of four $B^{3+}$ spins is antiparallel to the $Mn^{2+}$ spins with canting away from the $c$-axis and towards the opposite direction of $Mn^{2+}$ spins. In other words, one A-site spin and two neighboring B-site spins construct a magnetic triangle and a ferrimagnetic ordering phase emerges. The whole exchange energy on B sites will lower the full spin Hamiltonian of spinel, and the absolute value of spin Hamiltonian decreases by substituting $V^{3+}$ ion by $Mn^{3+}$ ion. Because the $J_{BB}^{ab}$ of $Mn^{3+}$-ion is two times larger than that of $V^{3+}$ ion [40], the transition temperature $T_2$ from CF to NCF of $V^{3+}$ occurs at higher temperature and the NCF of $Mn^{3+}$ at lower temperature $T_3$. Just as illustrated before, the second NCF phase was not captured by magnetization for $x \leq 0.4$ region due to lack of enough $Mn^{3+}$. For high doping region ($x \geq 0.5$), although $T_3$ occurs, it does not change significantly with $x$ changing.

As $Mn_{1-x}Co_xV_2O_4$ [20] and $Fe_{1-x}Co_xV_2O_4$ [43], it is also needed to be pointed out the decoupling of lattice and magnetic transitions in $Mn_{1+x}V_{2-x}O_4$ due to the induced competition between the orbital ordering and electronic itinerancy.

## 5. Conclusion

In summary, the orbital effects of $Mn^{3+}$ and $V^{3+}$ ions in $Mn_{1+x}V_{2-x}O_4$ have been studied. Both the Néel temperature $T_1$ and magnetic transition temperature from CF to NCF decreased with the Mn content increasing due to the competition of $J_{AB}$ and $J_{BB}$ and corresponding decreasing chemical pressure, and the decreasing canting angle of $V^{3+}$-$V^{3+}$ induced larger exchange energy on B sites. For $0.0 \leq x < 0.1$, the cubic to tetragonal ($c/a < 1$) phase transition was observed as $MnV_2O_4$; For $0.1 \leq x \leq 0.3$, the system showed a tetragonal phase with $c/a > 1$ below $T_S$ due to the orbital ordering of $Mn^{3+}$; For $x \geq 0.4$, the crystal structure keeps tetragonal phase up to room temperature. Except the CF-NCF transition on $V^{3+}$-ions, a $Mn^{3+}$-ions noncollinear magnetic ordering phase was observed at about 15 K for $x \geq 0.5$. Finally, the CF-NCF transition is decoupled from the cubic to tetragonal structural phase transition, which indicates there is no one-to-one corresponding relationship between these two transitions.

**Acknowledgments**

J.M. thanks helpful discussion with Dr. Q. Zhang, SNS, ORNL. J.J., W.W., G.W., Q.R. and J.M. are supported by the MOST and NSFC through projects with Grant Nos. 2016YFA0300501, 11774223 and U1732154. J.M. acknowledges additional support from a Shanghai talent program. G. L. is supported by China Postdoctoral Science Foundation with Grant No. 2019M661474. Research at the University of Tennessee is supported by the National Science Foundation, Division of Materials Research under award NSF-DMR-2003117 (H.Z., Q.H. and R. S.). H.P.Z. and M.Z.L. are supported by NSFC (No. 51631003). A portion of this research used resources at the High Flux Isotope Reactor and the Spallation Neutron Source, which are DOE Office of Science User Facilities operated by Oak Ridge National Laboratory.